\documentstyle[12pt]{ioplppt}
\begin{document}
\jl{6}

\title{Inflationary cosmology and thermodynamics}

\author{
Edgard Gunzig\dag\,,
Roy Maartens\ddag\ and
Alexei V. Nesteruk\ddag\ 
}

\address{\dag\
Faculte des
Sciences, Instituts Internationaux de Chimie et Physique Solvay,
Universit\'{e} Libre de Bruxelles,
1050~Bruxelles, Belgium}

\address{\ddag\ School of Computer Science and Mathematics,
University of Portsmouth,
Portsmouth~PO1~2EG, Britain}

\begin{abstract}

We present a simple and thermodynamically consistent
cosmology with a phenomenological model of quantum
creation of radiation due to vacuum decay.
Thermodynamics and Einstein's equations lead to an 
equation in which $H$ is determined by the particle
number $N$. The model is completed by specifying 
the particle creation
rate $\Gamma=\dot{N}/N$, which leads to a second-order evolution
equation for $H$. We propose a simple $\Gamma$ that is naturally
defined and that conforms to
the thermodynamical conditions: (a) the entropy production rate 
starts at a maximum; 
(b) the initial vacuum (for radiation) is a non-singular
regular  vacuum;
and (c) the creation
rate is initially higher than the expansion rate $H$, but then falls
below $H$.
The evolution equation for $H$ then
has a remarkably simple exact solution, in which
a non-adiabatic
inflationary era exits smoothly
to the radiation era, without a reheating transition.
For this solution, we give exact expressions
for the cosmic scale factor, energy density of radiation and vacuum,
temperature, entropy and super-horizon scalar perturbations.

\end{abstract}

\pacs{9880C, 9880H, 9530T, 0540}


\section{Introduction}

The creation of matter and entropy from
vacuum has been studied via quantum
field theory in curved spacetime
(see for example \cite{hupark77,creation}).
Most cosmological models exhibit a singularity which
presents  difficulties  for interpreting quantum effects, because
all  macroscopic parameters of created particles are infinite
there.  This leads to the  problem of the initial vacuum.
A regular vacuum for a species of created particles
can be defined in simple terms as a state  where all  mean
values describing the particles,
such as energy density, number density, entropy etc., are
zero.  But this simple condition  is not achieved in
many scenarios, so that either one has to postulate an initial
state beyond the singularity, or to assume that there was a nonzero
number of particles at the initial vacuum.

One attempt to
overcome these problems is via
incorporating the effect of particle creation
into Einstein's field equations. For example, in
the papers of the Brussels group \cite{edgard1},
the quantum effect of particle creation  is considered
in the context  of the thermodynamics of open systems,
where it is interpreted as an additional negative
pressure, which emerges from a re-interpretation of
the energy-momentum tensor.
This effect is irreversible in the sense that spacetime
can produce matter, leading to growth of entropy,
while the reverse process is thermodynamically
forbidden.

The main difference with our present paper is that
in \cite{edgard1} the law of massive particle creation, 
i.e. the mechanism
of energy flow from the gravitational field to matter, 
leads to a non-zero number of
particles at the beginning of expansion, described as a fluctuation
of the regular vacuum.  
These results were
recently generalized in  a covariant form in \cite{lima1}.
Our approach differs from that of \cite{edgard1,lima1}
in that we do not modify the field equations.
Instead, we interpret the source of created particles
as a decaying vacuum, described phenomenologically by a time-dependent
cosmological `constant' $\Lambda (t)$.

A number of decaying vacuum models has appeared in the literature
(see \cite{lima2,vl} and references cited there).
Inflationary models with fixed cosmological constant and cold dark
matter have been successful in accounting for the microwave
background and large-scale structure observations, while also
solving the age problem. However, these models are challenged by
the reduced upper limits on $\Lambda$ arising from the Supernova
Cosmology Project, and also by the long-standing problem of
reconciling the very large early-universe vacuum energy density
with the very low late-universe limits \cite{vl}. One resolution
of these problems is a decaying $\Lambda$.
In common with \cite{lima2}, we attempt to provide some
clear and consistent physical
motivation for the particular form of vacuum decay, rather than
an ad hoc prescription. In ad hoc prescriptions, the functional
form of $\Lambda(t)$ or $\Lambda(a)$ (where $a$ is the scale factor)
is effectively assumed a priori. Often power-law forms
for $\Lambda$ are assumed (see, for
example, \cite{new} and references cited there).
Exponential decay laws have also been assumed \cite{spindel}.
Typically, the solutions arising from ad hoc prescriptions
for $\Lambda$ are
rather complicated, and moreover, it is often difficult
to provide a consistent simple interpretation
of the features of particle creation, entropy and thermodynamics.

In contrast to many other models, we propose a simple, exact and
thermodynamically consistent cosmological history. The latter 
originates from {\em a regular initial vacuum with a maximal
initial entropy production rate.} Together with a naturally
defined creation rate, this leads to a simple expansion law and
thermodynamic properties, and to a definite estimate for the
total entropy in the universe. The very existence of an initial
maximal entropy production rate reflects a subtle interplay
between the conservation equation, the second law of thermodynamics
and Einstein's equations, which is at the heart of our model.

Non-adiabatic inflationary models
differ from the standard models (see for example
\cite{kolb&turner90}), in that: (a) radiation is created continuously
during inflation, rather than during reheating; (b) the
continuous vacuum decay itself
initiates a smooth exit from inflation to the radiation era; (c)
entropy and heat production take place continuously, without
the need for reheating. In the standard approach,
the scalar field drives
adiabatic (i.e., isentropic)
inflation, followed by a non-equilibrium reheating era
when the field decays into radiation and inflation is brought to an
end. The potential of the
field is then the key ingredient. In the alternative
approach, the key ingredient
is essentially the model of vacuum decay. In contrast to the
ad hoc models that assume functional forms for
$\Lambda$, and as an alternative to the thermodynamically motivated
model of \cite{lima2}, we arrive at a phenomenological model of decay
by imposing simple and thermodynamically
consistent physical requirements.
A related physically consistent model
can be found in the
`warm' inflationary scenario \cite{b}. 

The first law of thermodynamics for open systems (with particle
creation) and Einstein's equations lead to a first-order equation
for the expansion rate $H=\dot{a}/a$,
whose source term is determined by the particle number $N$.
A further equation in $H$ and $N$ arises from a simple model
for the particle creation rate $\Gamma=\dot{N}/N$. We impose
the thermodynamical non-equilibrium condition
that $\Gamma$, and hence the entropy production rate, is maximal
at the beginning of expansion.
A further initial condition is that the initial vacuum for
radiation is non-singular.
We also require that $\Gamma>H$
initially, so that created particles are thermalized, while
$\Gamma<H$ later on, as particle production becomes
insignificant. Our simple model for $\Gamma$ is naturally
defined by the gravitational dynamics and
conforms to these thermodynamical
requirements. We decouple the equations to get a second-order 
evolution for $H$, and we find a remarkably simple exact solution
$H(a)$.

Since the exit from inflation to the radiation era is smooth,
we avoid the problem of matching at the transition.
A similar
smooth evolution has been used in
\cite{roydavid,alexei_96,caldwell}, but
in the context of adiabatic
inflation, and without a consistent physical foundation.
In effect, we show that the ad hoc form of $H(a)$
given in \cite{roydavid} follows from our simple physical conditions
and thermodynamic arguments.
In \cite{b1}, a
kinematic analysis is given for various
non-adiabatic inflationary evolutions with smooth exit, but
these evolutions are outside the scope of our model.

The choice of $a$ as dynamical variable and the very simple form of
$H(a)$ that meets the physical conditions, lead
to elegant expressions for all parameters
describing the radiation and decaying vacuum, and also to a
physically transparent interpretation of these results, including
the estimate of entropy. In addition, the equation for
super-horizon scalar perturbations can be solved exactly for this
form of $H(a)$.

We present in  Sec. 2 the evolution equation for $H(a)$ and its
simple solution, that follow from our simple physical constraints.
In Sec. 3  we analyze the thermodynamics
of the radiation produced in the course of vacuum decay,
and estimate the entropy of the created radiation.
Sec. 4 contains a summary and concluding remarks.
In a subsequent paper, we
will discuss generalizations of the present model.

We use units with $8\pi G$, $c$ and $k_{\rm B}$ equal 1.


\section{The simple model}

Consider a spatially flat Friedmann-Lemaitre-Robertson-Walker
universe
\[
ds^2 = g_{\mu\nu}dx^\mu dx^\nu=
dt^2 - a^2(t) \left[dx^2 + dy^2 + dz^2\right]\,,
\]
containing matter with equation of state
\[
p = \gamma \rho,
\]
where $\gamma$ ($0\leq \gamma < 1$) is  a constant parameter.
Later we will specialize to the case of radiation, $\gamma={1\over3}$.
The energy-momentum tensor of matter is
\[
T^{\rm M}_{\mu\nu}= \rho(t)\left[ (\gamma +1) u_\mu u_\nu
+ \gamma g_{\mu\nu}\right]\,,~~u_\mu=\delta_\mu{}^0\,,
\]
while the energy-momentum tensor corresponding to
the quantum vacuum energy is
\[
T^{\rm Q}_{\mu\nu} \equiv \langle \widehat{T}^{\rm Q}_{\mu\nu}\rangle=
\Lambda (t)g_{\mu\nu}\,.
\]
Then the conservation equations
$\nabla^\nu(T^{\rm M}_{\mu\nu}+T^{\rm Q}_{\mu\nu})=0$
reduce to
\begin{equation}
\dot{\rho} +  3 (\gamma +1) H \rho = -
\dot{\Lambda}\,,
\label{conserv}
\end{equation}
showing how energy is transferred from the decaying vacuum to matter.
Note that (\ref{conserv}) is equivalent to the energy balance
of an imperfect fluid with scalar viscous pressure
\[
\Pi={\dot{\Lambda}\over 3H} \,.
\]
This is an example of the known result that cosmological particle
production may be interpreted as an effective bulk viscous
pressure (see \cite{z} and references cited there).

The field equations $R_{\mu\nu}-{1\over2}Rg_{\mu\nu}
=T^{\rm M}_{\mu\nu}+T^{\rm Q}_{\mu\nu}$ are
\begin{eqnarray}
3 H^2 &=& \rho + \Lambda \,, \label{einst0}\\
2 \dot{H} + 3 H^2 &=& - \gamma \rho  +  \Lambda \,,
\label{einst1}
\end{eqnarray}
and if both are satisfied then the energy equation (\ref{conserv})
follows identically.

Following \cite{roydavid}, we use $a$ as a dynamical
variable instead of $t$, and so we
consider  the Hubble rate as $H=H(a)$.
Then equations (\ref{einst0}) and (\ref{einst1})
give
\begin{eqnarray}
\rho (a) &=& -  \frac{2}{(\gamma +1)} \, a H(a) H'(a) \,,
\label{rho1}\\
\Lambda (a) &=&  3 H^2(a)  +  \frac{2}{(\gamma + 1)}
\, a  H(a)  H' (a)  \,,
\label{lambd1}
\end{eqnarray}
where primes denote $d/da$. Given $H(a)$ and $\gamma$, 
we can calculate $\rho(a)$ and $\Lambda(a)$.
In  order to determine further properties of  $H(a)$,
we will impose simple physical requirements.

We assume that the evolution of $\rho$  is governed not only
by expansion but also by  the creation of particles, i.e. from
the thermodynamic point of view,
we have an open system. According to \cite{edgard1}, the first
law of thermodynamics generalized for open systems is
\begin{equation}
d(\rho V) + p d V -  \left({\rho+p\over n}\right) d(nV) = 0 \,,
\label{firstlaw}
\end{equation}
where $p= \gamma \rho$ is the  pressure,
$n=N/V$ is the
particle number density, $N$ is the number of particles in the
observable universe,
and $V\propto a^3$ is the comoving volume of the observable
universe.
The thermodynamic equation (\ref{firstlaw}) implies
\[
{\left(\rho a^3\right)'\over \rho a^3}+\gamma{\left(a^3\right)' \over
a^3}-(\gamma+1){N'\over N}=0 \,,
\]
which integrates to
\begin{equation}
\rho={A\over(\gamma+1)}\left({N\over a^3}\right)^{\gamma+1} \,,
\label{numviahubble}
\end{equation}
where $A$ is a positive constant.
From now on, we take $\gamma={1\over3}$, i.e. we assume that only
radiation is created. This will also allow us
to define entropy production via the photon number.

By equation (\ref{rho1}), equation (\ref{numviahubble}) 
can be rewritten as an {\em evolution
equation for $H$, with source term determined by the particle
number:}
\begin{equation}
\frac{d}{d a } H^{2}(a) = -  A\, \frac{N^{4/3}(a)}
{a^5}\,.
\label{eqhubble}
\end{equation}
This is the fundamental equation which follows from Einstein's
equations and the thermodynamic equation (\ref{firstlaw}).
The creation mechanism is phenomenologically encoded in the
source term $N(a)$. 
The model must be completed via an equation that determines
$N$.
If the creation rate
of radiation $\Gamma=\dot{N}/N=aHN'/N$ is given, then equation
(\ref{eqhubble}) implies the second-order equation
\begin{equation}
3aHH''+3aH'^2+\left(15H-4\Gamma\right)H'=0 \,.
\label{-}\end{equation}
We seek a model in which 
most of the particle creation effectively takes place 
in the very early universe, starting from a regular vacuum.
More precisely, we 
impose the following thermodynamical 
requirements:\\ 
(a) Maximal
entropy production rate (equivalently, maximal particle creation
rate) at the beginning of expansion, so that the universe starts
in a state furthest away from equilibrium and then tends toward
equilibrium as the expansion proceeds. \\
(b) A true (regular) vacuum for radiation initially,
so that $\rho\rightarrow0$ as $a\rightarrow0$. \\
(c) $\Gamma>H$ in the very early universe, so that
we can treat the created radiation as forming a
thermalized heat bath. Subsequently, the
creation rate should fall behind the expansion rate as particle
creation becomes dynamically insignificant.

The fundamental physical quantities that are naturally defined 
by the gravitational dynamics in our model
are the expansion rate $H$ and the total
energy density $U=\rho+\Lambda$. Both of these quantities can
define in a natural way a gravitational
creation rate $\Gamma$. The
simplest model $\Gamma\propto H$ fails to satisfy requirement
(c) above. Furthermore, it leads to a solution
$H(a)$ of the evolution equation (\ref{-}) which violates
requirement (b).
Therefore we propose $\Gamma\propto U$. This will satisfy
requirement (a) if (b) holds, since the initial condition (b)
implies
that $U$ approaches its maximum value $U(0)=\Lambda(0)$ as 
$a\rightarrow0$. Below we show that the solution of equation
(\ref{-}) does verify condition (b).
The increase of $\rho$ due to creation in
the very early universe partially
offsets the decrease in $\Lambda$, so that $U$
decreases slowly in the very early universe, and the entropy
production rate remains high. 
Requirement (c) will be satisfied because 
in the radiation era, $\Lambda$ becomes negligible and
$\rho$ decays like $H^2$. Thus $\Gamma$ 
decreases more rapidly than $H$, and will 
have fallen below $H$ at some epoch.

Hence we propose the simple model that {\em the particle
creation rate is proportional to the total energy density.}
By the Friedmann equation (\ref{einst0}), it follows that 
\begin{equation}
\Gamma=\alpha H_{\rm e}\left({H\over H_{\rm e}}\right)^2 \,,
\label{cr}\end{equation}
where $\alpha$ is a dimensionless free parameter, and
$H_{\rm e}=H(a_{\rm e})$, where $a_{\rm e}$ is some fixed epoch.
Equation (\ref{-}) becomes
the
{\em decoupled, second-order evolution equation for $H(a)$:}
\begin{equation}
3aHH''+3aH'^2+15HH'-4{\alpha\over H_{\rm e}}H^2H'=0\,.
\label{roy}\end{equation}
This equation has the first integral
\begin{equation}
3aHH'+6H^2-{4\alpha\over 3H_{\rm e}}H^3=0 \,,
\label{cr2}\end{equation}
where we have used the fact that $H$ and $aHH'$ tend to zero
for large $a$ (i.e. in the radiation era), 
in order to remove a constant of integration.
The solution of equation (\ref{cr2}) is
\[
\beta\left({a\over a_{\rm e}}\right)=\left[
{9\over 2\alpha}\left({H_{\rm e}\over H}\right)-1\right]^{1/2}\,,
\]
where $\beta$ is a constant, and we have taken into account that
$H$ is a decreasing function. Evaluating at $a=a_{\rm e}$, we see
that
\begin{equation}
1+\beta^2={9\over2\alpha}\,,
\label{cr3}\end{equation}
which implies the constraint $\alpha\leq{9\over2}$ on the
creation parameter $\alpha$.
Re-arranging the solution, we
find the remarkably simple form for the 
expansion rate that follows from our thermodynamic model:
\begin{equation}
{H\over H_{\rm e}}=(1+\beta^2)\left[{a_{\rm e}^2\over
a_{\rm e}^2+\beta^2a^2}\right]\,.
\label{beta}\end{equation}

This solution approaches de Sitter inflation as $a\rightarrow0$,
i.e. $H\sim$ constant,
and it becomes radiation-like for $a\rightarrow\infty$, i.e.
$H\sim a^{-2}$. It follows that as $a\rightarrow0$, the cosmic
proper time $t\rightarrow-\infty$.
A naturally defined epoch is $a_{\rm ex}=a(t_{\rm ex})$ 
of exit from inflation, which
is defined by $\ddot{a}(t_{\rm ex}) = 0$, or equivalently
$H(a_{\rm ex}) = - a_{\rm ex} H'(a_{\rm ex})$. 
It follows from (\ref{beta}) that $a_{\rm e}=\beta a_{\rm ex}$.

These results reflect a subtle interplay between Einstein's
gravitational dynamics and thermodynamical constraints on
particle production. The decay of the vacuum
into radiation drives inflation, but the same decay, by reducing the 
vacuum energy density, leads to a deceleration of expansion and
a smooth exit from inflation. Since the initial vacuum for
radiation is regular, the particle number and hence entropy
are initially zero. As we show below, the total entropy produced
in the observable universe in an infinite time is finite. 
This is consistent with the existence of a smooth exit, since
unending inflation would produce infinite entropy.
The initial rate of entropy production is maximal, reflecting the
feature that the universe starts furthest from equilibrium 
and approaches asymptotically a state of equilibrium. We note
also that the initial entropy production rate is a finite
maximum value, rather than being unbounded from above. The latter
possibility is ruled out by the de Sitter-like nature of the
inflationary expansion, which implies that the expansion rate
$H$ is bounded from above (unlike power-law inflation for
example).

The freedom in $\beta$, or equivalently $\alpha$, by
equation (\ref{cr3}), provides us with an extra adjustable parameter.
However, for simplicity, we will not use this freedom, 
since the subsequent results are not modified in any essential way
for general $\beta$.
Henceforth we take $\beta=1$, i.e. $\alpha={9\over4}$,
which means that $a_{\rm e}=a_{\rm ex}$. 
Thus, we arrive finally at the simple expansion rate 
\begin{equation}
H(a) =
2H_{\rm e}\left({a_{\rm e}^2\over a_{\rm e}^2 + a^2}\right)\,.
\label{hubble}
\end{equation}
This form of $H(a)$
 was given in \cite{roydavid} as an ad hoc toy model
to achieve smooth exit from inflation to radiation, but without
a physical basis such as that given here.
The expression for the
cosmic proper time follows on integrating equation (\ref{hubble}):
\[
t=t_{\rm e}+{1\over 4H_{\rm e}}\left[
\ln\left({a\over a_{\rm e}}\right)^2+
\left({a\over a_{\rm e}}\right)^2-1\right]\,.
\]


\section{Thermodynamics of radiation}

On substituting equation (\ref{hubble}) into
equations (\ref{rho1}) and (\ref{lambd1}), we find exact
expressions for the energy density of radiation and the vacuum:
\begin{equation}
\rho (a) = 12H_{\rm e}^2\left({a\over
a_{\rm e}}\right)^2\left({a_{\rm e}^2\over a_{\rm e}^2
+a^2}\right)^3\,,~~~~~
\Lambda (a) = 12H_{\rm e}^2\left({a_{\rm e}^2\over
a_{\rm e}^2+a^2}\right)^3 \,.
\label{rholam}
\end{equation}
It follows that $\Lambda (0)  = 12 H_{\rm e}^{2}$. Note that
(\ref{hubble}) implies $H(0)=2H_{\rm e}$.
Note also that the effective bulk viscous pressure arising from
particle production has the form
\[
\Pi\equiv {\dot{\Lambda}\over 3H}=-\left({4\Gamma\over 9H}\right)\rho
=-\left({2a_{\rm e}^2\over a_{\rm e}^2+a^2}\right)\rho \,.
\]

Now $\rho$ reaches a maximum at $a_{\rm m} = a_{\rm e}/\sqrt{2}$, with
\[
\rho_{\rm m} \equiv \rho(a_{\rm m}) =
{\textstyle{16\over9}}\, H_{\rm e}^{2}\,,
~~~~~ \Lambda (a_{\rm m}) = 2 \rho_{\rm m} \,.
\]
Note also that $\rho$ and $\Lambda$ are equal at exit:
\[
\rho (a_{\rm e}) = \Lambda (a_{\rm e}) = {\textstyle{3\over2}}
H_{\rm e}^{2} \,,
\]
while for $a \gg a_{\rm e}$, i.e. during radiation-domination,
\[
\rho (a) \sim \frac{1}{a^4}\,,~~~~ \Lambda (a) \sim \frac{1}{a^6}\,,
\]
so that $\Lambda$ rapidly becomes negligible.

The formulas (\ref{rholam}) reflect the
creation of radiation due to vacuum decay.
The initial value $\rho (0) = 0$ confirms that the field
corresponding to radiation is initially in a regular vacuum state.
This means that the formula (\ref{rholam}) gives an
absolute measure of radiation produced in the universe, as a
result of conversion of energy from the vacuum described by
$\Lambda$.

Substituting equation (\ref{hubble})
into equation (\ref{numviahubble}) we get the exact form for the
particle number
\begin{equation}
N(a) = N_{\infty} \left(\frac{a^2}{a_{\rm e}^2 + a^2}\right)^{9/4}\,,
\label{number}
\end{equation}
where $N_{\infty} $ 
is usually taken to be about $10^{88}$.
It follows that about $2\times 10^{87}$ particles have been created
at exit. Note that the initial conditions and the evolution
equations in our model imply that {\em a finite number of
particles is produced in the observable universe
during the entire expansion.}

Since $N(0)= n(0) = 0$,
the initial state of the field has no particles, i.e. it is a
regular vacuum. The number density is
\begin{equation}
n(a) = 2^{9/4}n_{\rm e}\left({a_{\rm e}\over a}\right)^3
\left(\frac{a^2}{a_{\rm e}^2
+ a^2}\right)^{9/4}\,,
\label{n}\end{equation}
and reaches its
maximum also at $a_{\rm m}$.

The creation rate of radiation is given by equations (\ref{cr}) 
and (\ref{hubble}) as
\begin{equation}
\Gamma=9H_{\rm e}\left({a_{\rm e}^2\over
a_{\rm e}^2+a^2}\right)^2={9\over4}H_{\rm e}
\left({H\over H_{\rm e}}\right)^2\,.
\label{crate}\end{equation}
In order to define the radiation temperature (and then entropy), we
would like to invoke
the standard black-body relation. A justification of this is as
follows. From equation (\ref{crate}), it follows that the creation
rate $\Gamma (a)$ exceeds the expansion rate $H(a)$ for
\[
a<\sqrt{{\textstyle{7\over2}}}\, a_{\rm e} \,.
\]
Thus it is reasonable to treat the created radiation as forming
a thermalized heat bath in the initial stage of expansion.
For $a>\sqrt{7\over2}a_{\rm e}$, the creation rate falls behind the
expansion rate, and created particles will be out of equilibrium.
However, by this stage of the expansion, the 
energy density in newly
created particles is too small to disturb the effective
thermalization. Thus it seems reasonable to use the
black-body relation
for the radiation throughout the expansion, and to define the
temperature by
\begin{equation}
T(a) = \frac{1}{3}\frac{\rho(a)}{n(a)} =
{H_{\rm e}^2\over 2^{1/4}n_{\rm e}}
\left({a_{\rm e}\over a}\right)\left({a^2
\over a_{\rm e}^2+a^2}\right)^{3/4}\,,
\label{temperature}
\end{equation}
where we have used equations
(\ref{rholam}) and (\ref{n}).
At the initial radiation vacuum, it is clear that
$T(0)=0$, and $T$ increases  to its maximum
value at $a_{\rm m}$:
\[
T_{\rm m} \equiv T(a_{\rm m}) = \left({2\over27}\right)^{1/4}
{H_{\rm e}^2 \over n_{\rm e}}\,.
\]
This temperature may be thought of as analogous to the reheating
temperature in standard models.
During the radiation era, i.e. for $a \gg a_{\rm e}$,
\[
T \sim a^{-1}\,,
\]
in agreement with the standard result for free radiation in
an expanding universe.

The formulas for $\rho$ and $n$ can be presented in the thermodynamic
form
\begin{equation}
\rho = 24\left(\frac{ n_{\rm e}^4}
{H_{\rm e}^6}\right) T^{4}\,,~~~~~
n = 8\left(\frac{ n_{\rm e}^4}{H_{\rm e}^6}\right) T^{3}\,.
\label{tdform}
\end{equation}
Combining now the Gibbs equation
\[
T dS = d(\rho V) + p d V\,,
\]
with equation (\ref{firstlaw}), and using the definition
(\ref{temperature}) of $T$, we obtain the
entropy of radiation in the observable universe as
\begin{equation}
S (a) = 4 N (a)\,,
\label{entropy}
\end{equation}
so that $S(0)=0$ as expected.
This gives a reasonable value for the
entropy produced during the overall evolution of the universe:
\begin{equation}
S_\infty = 4 N_{\infty} \approx
4\times 10^{88}\,.
\label{entnumb}
\end{equation}
In the standard model (adiabatic inflation followed by reheating),
one can estimate the entropy production due to reheating by
matching exact de Sitter inflation to 
an exact radiation era, with an instantaneous transition 
at $a_{\rm e}$. This gives \cite{roydavid}:
\[
S_{\rm e} = {\textstyle{4\over3}}g^{1/4} \, \rho_{\rm e}^{3/4}=
S_\infty\,,
\]
where $g \sim 100$, and leads to a value
of the same order of magnitude as our result.

\section{ Conclusion}

We have considered a simple and thermodynamically consistent scenario
encompassing the decay of the vacuum, the creation of radiation
and entropy, and
a natural smooth transition from inflationary to radiation-dominated
expansion.
In order to treat all matter in the universe as created
from a regular vacuum, we impose the condition that
$\rho\rightarrow0$ as $a\rightarrow0$. 
We impose the further initial condition 
that the entropy production
rate is a maximum. A simple model for the particle creation rate
$\Gamma$,
i.e. that $\Gamma$ is proportional to the comoving total 
energy density,
is shown to be consistent with these initial conditions and
with the requirement $\Gamma$ should start above and then fall
below the expansion rate $H$.
We
showed that the field equations 
and the
first law of thermodynamics (\ref{firstlaw})
generalized for open systems  with creation of matter, 
then imply a second-order evolution equation (\ref{roy}) for $H(a)$.
This equation has the remarkably simple solution (\ref{beta}),
and we used this together with
black-body thermodynamics and the Gibbs equation to
define the temperature and entropy.

Our postulate $\Gamma\propto H^2$ for the creation rate, can be
contrasted with other work. For example, in \cite{edgard1}, 
$\dot{N}\propto H^2$, while $\Gamma$ is constant; in \cite{spindel},
$\Lambda\propto \exp(-t/\tau)$, and there is no simple expression for
$\Gamma$ in terms of $H$. The postulates in these papers are ad hoc,
whereas we have tried to give a thermodynamic justification for our 
postulate. It is also interesting to compare our 
decaying-vacuum/ radiation model with scalar-field/ radiation 
models \cite{b,or}. In these latter models, a phenomenological
term is introduced into the Klein-Gordon equation for the scalar
field $\phi$, describing the interaction between the decaying field
and radiation. The resulting energy balance equation is
\[
\dot{\rho}+4H\rho=\Gamma_\phi\dot{\phi}^2 \,,
\]
where $\Gamma_\phi$ is the phenomenological decay rate.
Comparing this with our energy balance (\ref{conserv}), and using
(\ref{numviahubble}), we see that $\Gamma_\phi\dot{\phi}^2$
corresponds to our ${4\over3}\Gamma \rho$. It is
not obvious whether a scalar field interpretation 
of our model exists that would produce the rate
$\Gamma_\phi\propto H^2$. This is a subject for further
investigation.

An important feature of the Hubble rate (\ref{hubble}) is that
$\lim_{a\rightarrow 0} H(a) = $ constant and
$\lim_{a\rightarrow 0} H' (a) = 0$.  According to
equations (\ref{rholam}) and (\ref{number}), this avoids any
divergences in $\rho$ and $N$ at $a=0$.
In addition,  it follows from the equations
(\ref{rholam}), (\ref{number}),
(\ref{temperature}) and (\ref{entropy})  that all thermodynamic
parameters describing the created
radiation vanish initially, i.e. all forms of matter
which we observe now in the universe were in a state of regular
vacuum, and all
constituents of the universe were created from this vacuum
in the course of the decay of the vacuum energy density
$\Lambda$ from its initial value $12H_{\rm e}^2$.
We also showed that our requirements imply finite particle 
and entropy
production during the entire expansion of the observable universe.

Our simple model leads to exact expressions for $t(a)$ and for
the thermodynamic variables. In addition,
the form (\ref{hubble}) of $H(a)$ leads to an
exact solution for the large-scale modes of
scalar perturbations, described gauge-invariantly by Bardeen's
potential $\Phi$. The solution is
\cite{roydaniel}:
\[
\Phi = C_+ \left(\frac{a^2}{a_{\rm e}^2 + a^2} \right) +
C_-\left({a_{\rm e}\over a}\right)\left({a_{\rm e}^2\over
a_{\rm e}^2 + a^2}\right)\,,
\]
where $C_\pm'=0$. We will not discuss the important question of
the source of these perturbations, except to mention that,
as pointed out in \cite{b}, non-adiabatic inflationary
scenarios allow for the possibility that the seeds of density
perturbations are of predominantly thermal, rather than quantum,
origin.

Further discussion of the physical processes taking place during
non-adiabatic inflation, and of their effects on baryogenesis,
the microwave background radiation and structure formation, is given
in \cite{b,b1,new,vl,cds}.

\newpage

\ack
This work was partially supported by EEC grants numbers
PSS* 0992 and CI1*-CT94-0004, and by OLAM - Fondation
pour la Recheche Fondamental, Brussels.

\end{document}